\documentclass[prd,showpacs,nofootinbib]{revtex4}
\usepackage[dvips]{graphicx}
\usepackage{amsmath}
\begin{document}
\title{Thermodynamic limit of the canonical partition function with
       respect to the quark number in QCD}
\author{Kenji Fukushima}
\email[E-mail: ]{fuku@nt1.c.u-tokyo.ac.jp}
\affiliation{Institute of Physics, University of Tokyo, 3-8-1 Komaba,
             Meguro-ku, Tokyo 153-8902, Japan}
\thanks{The present address: Department of Physics, University of
        Tokyo, 7-3-1 Hongo, Bunkyo-ku, Tokyo 113-0033, Japan}
\begin{abstract}
We investigate QCD in the canonical ensemble with respect to the quark
number. We reveal that the canonical description in which the quark
number is fixed would be reduced to the grand canonical description
under the thermodynamic limit. Since the grand canonical ensemble
contains fluctuations of the quark number, the idea of the canonical
ensemble is of no use for the purpose of defining order parameters for
the deconfinement transition. We clarify the origin of such reduction
and propose an idea to define the order parameter. The Monte-Carlo
simulation by means of the spin system, which is an effective model of
QCD at finite temperature, shows prosperous behavior though the
results suffer from the severe statistical error due to the sign
problem.
\end{abstract}
\maketitle


\section{INTRODUCTION}
Extensive efforts have been shedding light upon the phase structure of
Quantum Chromodynamics (QCD) at finite temperature, though there
remain many subtleties yet. The center symmetry is a prosperous
implement to characterize the deconfinement transition in pure
non-abelian gauge theories (gauge theories without dynamical
quarks)~\cite{sve82}. The order parameter to examine whether the
symmetry is broken or not is the expectation value of the Polyakov
loop~\cite{pol78}, that is, the Wilson line winding around the
Euclidean thermal torus. The Polyakov loop vanishes in the confined
phase, while it takes a finite value in the deconfined phase. In pure
gauge theories the critical properties associated with the
deconfinement transition have been ascertained by the lattice
Monte-Carlo simulations. The results are in accord with those
anticipated from the center symmetry and the universality
argument~\cite{mcl81}.

In contrast to pure gauge theories any definite indicator in order to
distinguish the deconfined phase from confined one is not established
so far yet for systems including dynamical quarks in the fundamental
representation. Once the thermal excitation of light quarks is
allowed, the center symmetry is broken explicitly. If we construct a
3-d effective model in terms of the order parameter, namely the
Polyakov loop, in the presence of dynamical quarks, those quark
contributions bring about an external magnetic-like field acting onto
the Polyakov loop, which breaks the center symmetry~\cite{ban83}. Then
it is obvious that the Polyakov loop is no longer a proper indicator
for confinement because it remains finite even in the confined phase
as well as in the deconfined phase due to the absence of the center
symmetry.

We note, however, that the notion of confinement should be still
articulate even when dynamical quarks are present simply because it is
an experimental fact. To find out an appropriate indicator to identify
the confined phase in the presence of dynamical quarks it is essential
to clarify where the dynamical screening against the Polyakov loop
should originate from. In fact, DeTar and McLerran~\cite{det82} argued
that some fractional excitations of dynamical quarks are responsible
for the explicit breaking of the center symmetry and proposed a new
order parameter for the deconfinement transition based on the
canonical description with respect to the quark triality or quark
number. This idea had been deserted since Meyer-Ortmanns~\cite{mey84}
showed that DeTar--McLerran's order parameter behaves unexpectedly and
always indicates the confining character for the Z(2) Higgs model on
the lattice. Although Meyer-Ortmanns has explicitly demonstrated a
failure in the proposed order parameter in the model study, it is not
apparent in principle what would cause DeTar--McLerran's formulation
not to fulfill the naive expectation as a proper order parameter.
Similar ideas for the deconfinement order parameter based on the
canonical ensemble have been proposed and discussed by several authors
also~\cite{wei87,ole92,fab95} but those arguments are essentially
integrated into DeTar--McLerran's first insight.

Interestingly enough, we can find a clear-sighted comment in
Meyer-Ortmanns' paper as follows: ``\textit{The states at zero
temperature have a finite particle number hence zero particle density,
while the relevant states at finite temperature have finite particle
density. Therefore one cannot check whether there exist states at zero
temperature with fractional baryon number by looking at what states
contribute to the partition function at finite temperature.}'' In her
analysis on the Z(2) Higgs model, we can discern no clear embodiment
for the statement quoted above. Nevertheless, we would emphasize that
this point is actually the most critical in constructing the order
parameter for the deconfinement transition. The purpose of the present
paper is to give a simple and transparent demonstration to disclose
what becomes of the partition function with the particle number kept
fixed under the thermodynamic limit.

The problem here can be also stated in more general grounds. The
question is whether the grand canonical description with zero quark
chemical potential is equivalent to the canonical description with
zero quark number or not. We will find that the canonical description
with zero quark \textit{number} would amount to the canonical
description with zero quark \textit{density} whenever the
thermodynamic limit is taken. As a result, DeTar--McLerran's idea to
project out the fractional excitations of dynamical quarks does not
work because the states with zero particle density may have any
fractional excitation of particles which becomes irrelevant eventually
in the thermodynamic limit. Then by all means can we reach the desired
canonical ensemble in which the particle number is fixed at zero? We
will propose one possibility in the present work.

This paper is organized as follows: In Sec.~\ref{sec:canonical} we
review DeTar--McLerran's idea from the point of view of the canonical
ensemble with respect to the quark number. Sec.~\ref{sec:limit} is
devoted to investigating the thermodynamic limit of the canonical
partition function. Using the simplest example we elucidate
analytically and numerically that the canonical partition function
would be reduced to the conventional grand canonical one under the
thermodynamic limit. We reveal the cause of the failure in defining an
order parameter for the deconfinement transition. In
Sec.~\ref{sec:idea}, we propose an idea to overcome the problem in
connection with taking the thermodynamic limit. Performing the
Monte-Carlo simulation, we put the idea to the test to confirm that
our order parameter would be prosperous. We also find that the problem
of constructing an order parameter for the deconfinement transition is
deeply related to the sign problem of the Dirac determinant. The
results of the numerical simulations suffer from the severe
statistical errors due to the sign problem. The concluding remarks are
in Sec.~\ref{sec:remarks}.


\section{IDEA OF THE CANONICAL ENSEMBLE}
\label{sec:canonical}


\subsection{Canonical ensemble with respect to the quark number}
We will briefly review the idea of the canonical ensemble with
emphasis on the possibility to define an order parameter for the
deconfinement transition.

\begin{figure}
\includegraphics[width=11cm]{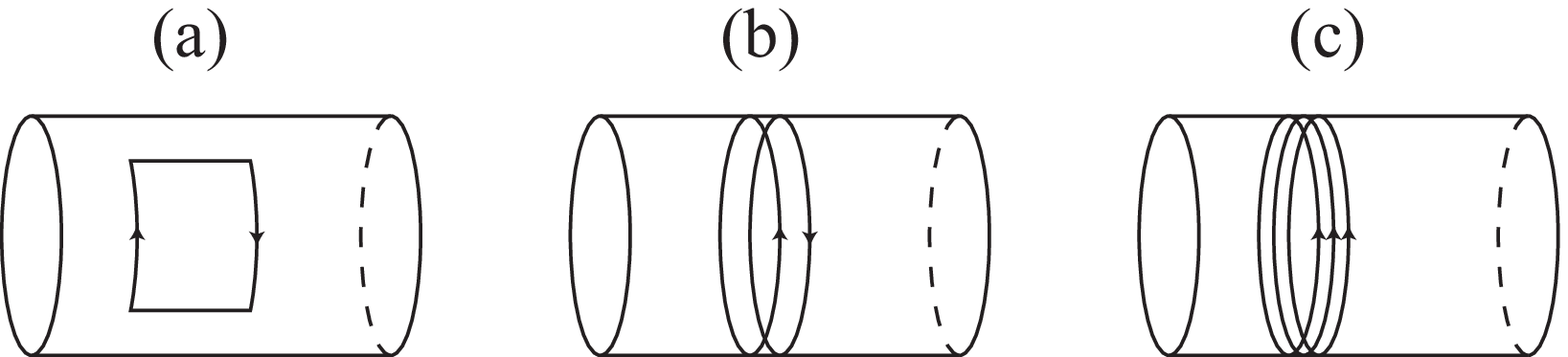}
\caption{Typical examples of the thermal excitations of dynamical
quarks which respect the center symmetry. The torus represents the
thermal $S^1$ and the spatial $R^3$. (a) Creation and annihilation
process of a quark ($q$) and an antiquark ($\bar{q}$). (b) Mesonic
excitation composed of $q\bar{q}$. (c) Baryonic excitation composed of
$qqq$ in the case of $N_{\mathrm{c}}=3$.}
\label{fig:ex_cs}
\includegraphics[width=11cm]{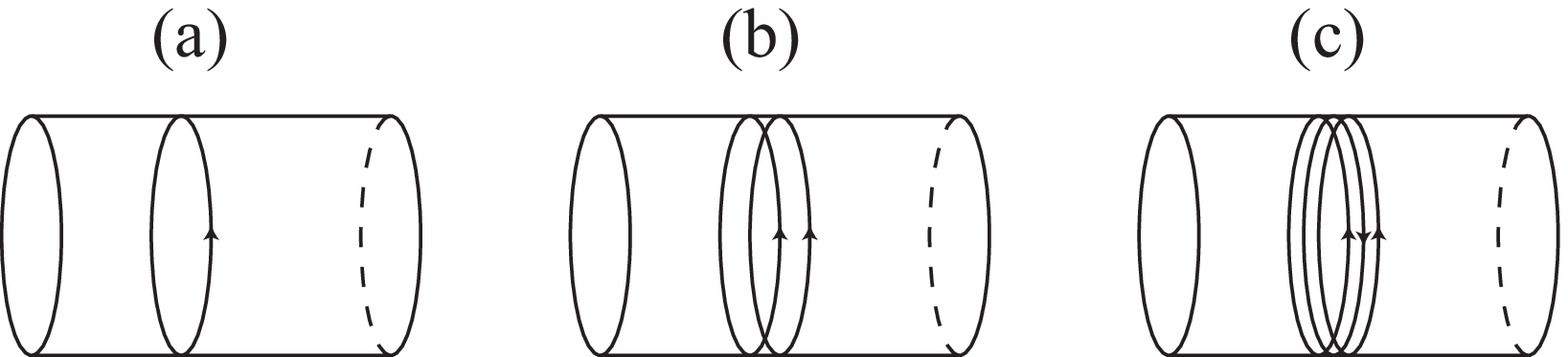}
\caption{Typical examples of the thermal excitations of dynamical
quarks which break the center symmetry. (a) Single quark ($q$)
excitation. (b) $qq$ excitation. (c) Mesonic and single quark
($q\bar{q}q$) excitation.}
\label{fig:ex_ns}
\end{figure}

First of all, we should notice that the conventional QCD partition
function at finite temperature and zero density is implicitly
formulated in the grand canonical ensemble with zero quark chemical
potential, that is equivalent to the canonical ensemble with zero
quark density under the thermodynamic limit. This fact is easily
understood as follows. In the imaginary-time formalism of the
finite-temperature field theories, the temporal (thermal) extent is
compactified to the inverse temperature. Then, in addition to the pair
excitations of a quark and an antiquark as depicted in
Fig.~\ref{fig:ex_cs}~(a), isolated quark excitations as shown in
Fig.~\ref{fig:ex_ns}~(a) also satisfy the quark current conservation
and thus the quark number can fluctuate thermally. In the
thermodynamic limit, however, the quark number conservation is
recovered because only the states with zero quark number dominate the
partition function at vanishing chemical potential. This point of view
has already been discussed in Ref.~\cite{azc98} and it seems
reasonable from the principle of the statistical mechanics.

Usually it does not matter in calculating physical observables whether
such fractional excitations of dynamical quarks are present. The
situation may be totally different for the manifestation of the center
symmetry. In the language of the lattice gauge theory, the center
symmetry is defined as the invariance under the center transformation,
that is, the transformation from the temporal link variables
$U_4(x_4=N_\tau)$ on a time slice altogether into
$z\cdot U_4(x_4=N_\tau)$, where $z$ is an element of the center of the
gauge group (Z(3) for the SU(3) gauge theory). For the quark thermal
contributions the Z(3) factor $z$ is multiplied every time dynamical
quarks wind around the thermal torus. Thus thermal excitations
gathered typically in Fig.~\ref{fig:ex_cs} are center symmetric
because $z\cdot z^\dagger=z^3=1$ for $z\in Z(3)$. Thermal excitations
shown in Fig.~\ref{fig:ex_ns} are, on the other hand, typical examples
which break the center symmetry. Then, what happens if we exclude any
thermal excitation like those in Fig.~\ref{fig:ex_ns} by using the
canonical description where the quark number is fixed? Once we employ
the canonical ensemble with zero quark number, for example, only such
excitations as shown in Fig.~\ref{fig:ex_cs}~(a)~(b) are allowed so
that the canonical partition function would regain the center
symmetry. Certainly the symmetry itself is explicitly broken in the
Lagrangian and nevertheless the physical states remain center
symmetric. Owing to the recovered symmetry, we can presume that the
Polyakov loop will provide a criterion for the deconfinement
transition in the same way as in the case of pure gauge theories.
Essentially, this is the idea proposed originally by DeTar and
McLerran in Ref.~\cite{det82}.

The canonical description of QCD at finite density is available by the
Legendre transformation from the quark chemical potential $\mu$ to the
quark number $n_q$~\cite{eng99}. (In this paper we simply write the
\textit{quark number} to mean the quark number minus the antiquark
number.) For later convenience, we explain the transformation
procedures in some mathematical details. To make our calculation
tangible, we adopt the Wilson fermion on the lattice from now on. The
final results, of course, can be understood not only in a specific
formalism but in a general way also.

The quark number density is the fourth component of the conserved
current given by
\begin{equation}
 j_4(x)=\kappa\{\bar{\psi}(x)(1-\gamma_4)U_4(x)\psi(x+\hat{4})-
  \bar{\psi}(x+\hat{4})(1+\gamma_4)U_4^\dagger(x)\psi(x)\},
\end{equation}
where $\kappa$ is the hopping parameter and the Wilson parameter is
chosen as $r=1$ for the sake of simplicity. By imposing the constraint
onto the configurations we can write the canonical partition function
with respect to the quark number $n_q$ as
\begin{align}
 Z_{\text{CE}}(n_q) &=\int\mathcal{D}U\mathcal{D}\bar{\psi}\mathcal{D}
  \psi\,\mathrm{e}^{-S_{\text{G}}[U]-S_{\text{F}}[U,\bar{\psi},\psi]}
  \,\delta\biggl(\frac{1}{N_\tau}\sum_x j_4(x)-n_q\biggr)\notag\\
 &=\int_0^{2\pi}\frac{\mathrm{d}\phi}{2\pi}\,\mathrm{e}^{-\mathrm{i}
  n_q\phi}\int\mathcal{D}U\mathcal{D}\bar{\psi}\mathcal{D}\psi\,
  \mathrm{e}^{-S_{\text{G}}[U]-S_{\text{F}}[U,\bar{\psi},\psi]+
  \mathrm{i}\frac{\phi a}{\beta}\sum j_4(x)},
\label{eq:CE_QCD}
\end{align}
where $S_{\text{G}}[U]$ is the gluon action and
$S_{\text{F}}[U,\bar{\psi},\psi]$ is the action of the Wilson fermion
with $r=1$. The lattice spacing $a$ and the number of lattices along
the temporal direction $N_\tau$ give the inverse temperature,
\textit{i.e.}, $\beta=N_\tau a$. The quark parts of the action are put
together as
\begin{align}
 &-S_{\text{F}}[U,\bar{\psi},\psi]+\mathrm{i}\frac{\phi a}{\beta}
  \sum_x j_4(x)\notag\\
 =&-\sum_x\biggl\{\bar{\psi}(x)\psi(x)-\kappa\sum_j\bigl[\bar{\psi}(x)
  (1-\gamma_j)U_j(x)\psi(x+\hat{j})+\bar{\psi}(x+\hat{j})(1+\gamma_j)
  U_j^\dagger(x)\psi(x)\bigl]\notag\\
 &-\kappa\biggl[\Bigl(1\!+\!\mathrm{i}\frac{\phi a}{\beta}\Bigr)
  \bar{\psi}(x)(1\!-\!\gamma_4)U_4(x)\psi(x\!+\!\hat{4})+\Bigl(1\!-\!
  \mathrm{i}\frac{\phi a}{\beta}\Bigr)\bar{\psi}(x\!+\!\hat{4})
  (1\!+\!\gamma_4)U_4^\dagger(x)\psi(x)\biggr]\biggr\}.
\end{align}
Here $\mathrm{i}\phi$ can be regarded as an imaginary chemical
potential and the integration over $\phi$ corresponds to the Legendre
transformation from the chemical potential to the quark
number~\cite{eng99}. $\hat{j}$ runs from $\hat{1}$ to $\hat{3}$ in the
spatial directions. In the same way as the usual prescription to treat
the chemical potential in the lattice gauge theories~\cite{has83} we
exponentiate the imaginary chemical potential as follows;
\begin{equation}
 1+\mathrm{i}\frac{\phi a}{\beta}\sim\exp\Bigl(\mathrm{i}
  \frac{\phi a}{\beta}\Bigr),\qquad
 1-\mathrm{i}\frac{\phi a}{\beta}\sim\exp\Bigl(-\mathrm{i}
  \frac{\phi a}{\beta}\Bigr),
\end{equation}
which makes only higher order corrections of $\mathrm{O}(a^2)$. The
fundamental reason to demand the above prescription is because of the
gauge invariance, or the center symmetry in the present case. With these
alterations the partition function given by Eq.~(\ref{eq:CE_QCD})
becomes exactly center symmetric for $n_q=0,\pm3,\pm6,\dots (0\mod3)$
as it should be. In order to see it apparently, transforming the quark
fields by
\begin{equation}
 \psi'(x)=\mathrm{e}^{\mathrm{i}\frac{\phi a}{\beta}x_4}\psi(x),
  \qquad \bar{\psi}'(x)=\mathrm{e}^{-\mathrm{i}\frac{\phi a}{\beta}
  x_4}\bar{\psi}(x),
\end{equation}
we rewrite the quark action in terms of the transformed fields as
\cite{eng99}
\begin{align}
 &\!-\tilde{S}_{\text{F}}[U,\bar{\psi}',\psi']\notag\\
 &\!+\kappa\!\sum_{\vec{x}}\Bigl[\mathrm{e}^{\mathrm{i}\phi}
  \bar{\psi}'(\vec{x},N_\tau)(1\!-\!\gamma_4)U_4(\vec{x},N_\tau)
  \psi'(\vec{x},1)\!+\!\mathrm{e}^{-\mathrm{i}\phi}\bar{\psi}'
  (\vec{x},1)(1\!+\!\gamma_4)U_4^\dagger(\vec{x},N_\tau)\psi'
  (\vec{x},N_\tau)\Bigr].
\label{eq:act_quark}
\end{align}
$\tilde{S}_{\text{F}}[U,\bar{\psi},\psi]$ is the rest of the action
from which the terms involving $U_4(\vec{x},N_\tau)$ are subtracted.
It is obvious that the center transformation $U_4(\vec{x},N_\tau)\to
\mathrm{e}^{\mathrm{i}2\pi k/3}  U_4(\vec{x},N_\tau)$ is equivalent to
the shift of $\phi$ by $2\pi k/3$. We can immediately recognize from
Eq.~(\ref{eq:CE_QCD}) that the canonical partition function is
certainly invariant under $\phi\to\phi+2\pi k/3$ as long as the quark
number satisfies $n_q=0\mod3$.

The integration of the partition function in terms of $\phi$ exactly
corresponds to the decomposition of the Dirac determinant into each
part with fixed quark winding number. This is seen transparently if
the Dirac determinant obtained from the quark action
Eq.~(\ref{eq:act_quark}), that is denoted by $\det D[U,\phi]$ from now
on, is calculated by means of the hopping parameter expansion. The
expanded terms are represented by closed quark paths similarly as
depicted in Figs.~\ref{fig:ex_cs} and \ref{fig:ex_ns}. Every time the
quark path winds around the thermal torus, it picks up the factor
$\mathrm{e}^{\mathrm{i}\phi}$ so that the Dirac determinant can be
decomposed according to the quark number $n_q$;
\begin{equation}
 \det D[U,\phi]=\sum_{n_q=-\infty}^\infty
  \mathrm{e}^{\mathrm{i}n_q\phi}\det D^{(n_q)}[U],
\label{eq:decom_det}
\end{equation}
where $\det D^{(n_q)}[U]$ is the part of the Dirac determinant with
the quark path winding $n_q$ times around the thermal torus. Thus the
integration over $\phi$ in Eq.~(\ref{eq:CE_QCD}) singles out some
specific combinations of the operators coming from the Dirac
determinant under the condition that the quark winding number is
fixed. This result is not particular to the case of the Wilson fermion
formalism on the lattice. In general, the canonical partition function
is given by
\begin{equation}
 Z_{\text{CE}}(n_q)=\int_0^{2\pi}\frac{\mathrm{d}\phi}{2\pi}\,
  \mathrm{e}^{-\mathrm{i}n_q\phi}\int\mathcal{D}U\det D[U,\phi]\,
  \mathrm{e}^{-S_{\text{G}}[U]}
\end{equation}
with the Dirac determinant $\det D[U,\phi]$ under the twisted boundary
conditions,
\begin{equation}
 \psi(\vec{x},x_4=\beta)=
  -\mathrm{e}^{\mathrm{i}\phi}\psi(\vec{x},x_4=0),\qquad
 \bar{\psi}(\vec{x},x_4=\beta)=
  -\mathrm{e}^{-\mathrm{i}\phi}\bar{\psi}(\vec{x},x_4=0).
\end{equation}
It leads to the canonical partition function  in the same sense as in
Ref.~\cite{ole92}, that is, the partition function with the Dirac
determinant replaced by its specific part with some fixed quark
excitations. It is interesting that the constraint onto the
configurations, imposed by Dirac's delta function, results in the
restriction onto the combination of the operators after all. This is
because of the non-local character of the quark (fermion) fields.


\subsection{DeTar--McLerran's order parameter}
Here we will explicitly construct DeTar--McLerran's order parameter
for the deconfinement transition. They introduced the quark triality
$t_q$ defined as $n_q\mod3$ and considered the canonical partition
function with respect to $t_q$. It is available from the canonical
partition function with respect to $n_q$ as
\begin{align}
 \tilde{Z}_{\text{CE}}(t_q)
  &=\sum_{m=-\infty}^\infty Z_{\text{CE}}(n_q=3m+t_q)\notag\\
  &=\frac{1}{3}\sum_{\phi=0,\pm2\pi/3}\mathrm{e}^{-\mathrm{i}t_q\phi}
   \int\mathcal{D}U\det D[U,\phi]\,\mathrm{e}^{-S_{\text{G}}[U]}.
\end{align}
Then DeTar--McLerran's order parameter is given by
\begin{equation}
 \mathcal{\tilde{P}_N}=\frac{\tilde{Z}_{\text{CE}}(t_q=\mathcal{N})}
  {\sum_{t_q}\tilde{Z}_{\text{CE}}(t_q)}
\label{eq:detar}
\end{equation}
for arbitrary $\mathcal{N}\neq0$. When the quark number is fixed
instead of the quark triality, we can define also in the same way as
\begin{equation}
 \mathcal{P_N}=\frac{Z_{\text{CE}}(n_q=\mathcal{N})}
  {\sum_{n_q}Z_{\text{CE}}(n_q)},
\end{equation}
which is expected to behave similarly to $\mathcal{\tilde{P}_N}$. In
the confined phase where all the physical excitations are
color-neutral, any thermal excitation of isolated quarks would cost
infinitely large energy. Therefore it follows that
$\tilde{Z}_{\text{CE}}(t_q=\mathcal{N})=Z_{\text{CE}}(n_q=\mathcal{N})
=0$ for $\mathcal{N}\neq0\mod3$. After the dynamical breaking of the
center symmetry occurs, fractional quark excitations are allowed at
finite energies so that the canonical partition function can take a
non-vanishing value. Hence, $\mathcal{\tilde{P}_N}$ and
$\mathcal{P_N}$ are anticipated to serve as order parameters for the
deconfinement transition.

We can manipulate the order parameter in other forms comparable with
the conventional definition, namely the Polyakov loop. As discussed in
Refs.~\cite{ole92,fab95}, the expectation value of the Polyakov loop
itself can be expected to be a proper order parameter in the canonical
ensemble. Because the center symmetry becomes manifest in the
canonical partition function with the quark number $n_q=0\mod3$ (or
the quark triality $t_q=0$), the deconfinement transition could be
characterized by the dynamical breaking of the center symmetry. Thus
another candidate for an order parameter for the deconfinement
transition is given by
$\langle\widetilde{\mathrm{tr}L(\vec{x})}\rangle_{\text{CE}}$ with
$t_q=0$, that is,
\begin{equation}
 \langle\widetilde{\mathrm{tr}L(\vec{x})}\rangle_{\text{CE}}=\frac{
  \frac{1}{3}\sum_\phi \int\mathcal{D}U\det D[U,\phi]\,\mathrm{e}^{
  -S_{\text{G}}[U]}\,\mathrm{tr}L(\vec{x})}{\tilde{Z}_{\text{CE}}
  (t_q=0)}.
\label{eq:pol}
\end{equation}
In the same way as in the case of $\mathcal{\tilde{P}_N}$ a similar
variation is also possible for
$\langle\widetilde{\mathrm{tr}L(\vec{x})}\rangle_{\text{CE}}$,
\textit{i.e.}, we can consider
$\langle\mathrm{tr}L(\vec{x})\rangle_{\text{CE}}$ with $n_q=0$ for
example as
\begin{equation}
 \langle\mathrm{tr}L(\vec{x})\rangle_{\text{CE}}=\frac{\int_0^{2\pi}
  \frac{\mathrm{d}\phi}{2\pi}\int\mathcal{D}U\det D[U,\phi]\,
  \mathrm{e}^{-S_{\text{G}}[U]}\,\mathrm{tr}L(\vec{x})}
  {Z_{\text{CE}}(n_q=0)},
\end{equation}
where $L(\vec{x})$ stands for the Polyakov loop defined by
\begin{equation}
 L(\vec{x})=\mathcal{T}\exp\biggl[-\mathrm{i}g\int_0^\beta
  \mathrm{d}x_4\,A_4(x_4)\biggr].
\end{equation}
Here $\mathcal{T}$ represents the time-ordering and $A_4(x)$ is the
temporal component of gauge fields.

We can explicitly evaluate the integration with respect to $\phi$ once
the actual form of $\det D[U,\phi]$ is given. Now we will exploit the
hopping parameter expansion. Following the calculation of
Ref.~\cite{gre84} the Dirac determinant amended by $\phi$ can be
readily expanded in the lowest order of the hopping parameter
expansion as follows;
\begin{equation}
 \det D[U,\phi]=\exp\biggl[H\sum_{\vec{x}}\bigl\{\mathrm{e}^{
  \mathrm{i}\phi}\mathrm{tr}L(\vec{x})+\mathrm{e}^{-\mathrm{i}\phi}
  \mathrm{tr}L^\dagger(\vec{x})\bigr\}\biggr],
\end{equation}
where $H=2(2\kappa)^{N_\tau}$ and $L(\vec{x})$ here is the Polyakov
loop defined on the lattice by
\begin{equation}
 L(\vec{x})=\prod_{x_4=a}^{N_\tau a}U_4(\vec{x},x_4),
\end{equation}
which can be graphically represented like in Fig.~\ref{fig:ex_ns}~(a).
With the notations $M=\sum\mathrm{tr}L(\vec{x})=|M|\,
\mathrm{e}^{\mathrm{i}\theta_M}$, the integration in terms of $\phi$
in Eq.~(\ref{eq:CE_QCD}) results in~\cite{mil87}
\begin{equation}
 \int_0^{2\pi}\frac{\mathrm{d}\phi}{2\pi}\,\mathrm{e}^{-\mathrm{i}
  n_q\phi}\cdot\mathrm{e}^{H(\mathrm{e}^{\mathrm{i}\phi}M+
  \mathrm{e}^{-\mathrm{i}\phi}M^\dagger)}
  =\mathrm{e}^{\mathrm{i}n_q\theta_M}I_{n_q}(2H|M|),
\label{eq:integ_phi}
\end{equation}
where $I_\nu(x)$ is the modified Bessel function of the first kind. By
differentiating the integrand with respect to $\phi$, we can have the
expression for the expectation value of the quark number as
\begin{equation}
 n_q =\langle HM\rangle_{n_q-1}-\langle HM^\dagger\rangle_{n_q+1}.
\label{eq:nq}
\end{equation}
This relation can be confirmed analytically also by the formula
$I_{\nu-1}(x)-I_{\nu+1}(x)=(2\nu/x)I_\nu(x)$. Also it can be
expressed as $n_q= 2\mathrm{i}H\mathrm{Im}\langle\sum\mathrm{tr}
L(\vec{x})\mathrm{e}^{\mathrm{i}\phi}\rangle_{n_q}$. This form of the
relation has been found already in Ref.~\cite{mil87}. It would be
worth mentioning that a different but similar relation has been
discussed also in a numerical way in Ref.~\cite{for00}, which suggests
that the above relation could persist beyound the leading order of the
hopping parameter expansion.

In the lowest order of the hopping parameter expansion, we can write
down DeTar--McLerran's order parameter as
\begin{equation}
 \mathcal{P_N}=\frac{\int\mathcal{D}U\,\mathrm{e}^{-S_{\text{G}}[U]
  +\mathrm{i}\mathcal{N}\theta_M}I_{\mathcal{N}}(2H|M|)}
  {\int\mathcal{D}U\,\mathrm{e}^{-S_{\text{G}}[U]+2H|M|\cos
  \theta_M}}.
\label{eq:order1}
\end{equation}
As for the expectation value of the Polyakov loop in the canonical
ensemble, we can immediately write down as
\begin{equation}
 \langle \mathrm{tr}L\rangle_{\text{CE}}=\frac{1}{V}\frac{\int
  \mathcal{D}U\,\mathrm{e}^{-S_{\text{G}}[U]+\mathrm{i}\theta_M}
  I_0(2H|M|)\,|M|}
  {\int\mathcal{D}U\,\mathrm{e}^{-S_{\text{G}}[U]}I_0(2H|M|)}.
\label{eq:order2}
\end{equation}

In the next section, we address the problem of taking the
thermodynamic limit for these would-be order parameters
(\ref{eq:order1}) and (\ref{eq:order2}). The point will turn out that
the sum of the Polyakov loop $M=\sum\mathrm{tr}L(\vec{x})$ is an
extensive quantity and as large as proportional to the volume.


\section{THERMODYNAMIC LIMIT}
\label{sec:limit}


\subsection{Failure as order parameters}
\begin{figure}
\includegraphics[width=9cm]{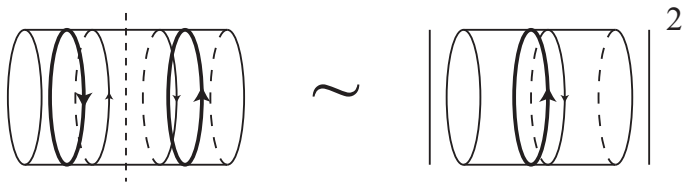}
\caption{The schematic picture for the emergence of fractional quark
excitations allowed even in the canonical ensemble under the
thermodynamic limit.}
\label{fig:emergence}
\end{figure}

We consider the thermodynamic limit here, that is, we make the system
volume go to infinity. Then we will find that both $\mathcal{P_N}$ and
$\langle\mathrm{tr}L\rangle_{\text{CE}}$ fail serving as order
parameters for the deconfinement transition. Roughly speaking, the
failure as order parameters stems from fractional quark excitations
permitted even in the canonical ensemble under the infinite volume
limit. As schematically depicted in Fig.~\ref{fig:emergence}, a pair
of a quark and an antiquark can be excited (drawn by the thin curves)
because such an excitation is mesonic and the net quark number is
zero. When the distance between source quarks (drawn by the thick
curves) gets large, it would be regarded as the system composed of two
mesonic excitations as shown in the left figure of
Fig.~\ref{fig:emergence}. In the limit of the infinite distance
between source quarks, two mesonic excitations should become
individual (clustering decomposing property as shown in the right
figure of Fig.~\ref{fig:emergence}), which means that the Polyakov
loop associated with a single quark excitation can be effectively
screened by the fractional excitation of dynamical quarks. Although
any fractional excitation of dynamical quarks is prohibited at first
by the definition of the canonical ensemble, it would be allowed
eventually in the infinitely large volume limit due to the clustering
decomposing property. As a result, $\mathcal{P}_{\mathcal{N}\neq0}$
amounts to non-zero $\mathcal{P}_{\mathcal{N}=0}$ and
$\langle\mathrm{tr}L\rangle_{\text{CE}}$ is reduced to the Polyakov
loop expectation value in the conventional grand canonical ensemble
under the thermodynamic limit.

Let us confirm the above intuitive understanding in the actual
expressions given by Eqs.~(\ref{eq:order1}) and (\ref{eq:order2}).
Without any spontaneous breaking of the center symmetry,
$\mathcal{P_N}$ is expected to vanish for $\mathcal{N}\neq0\mod3$.
This expected property derives from the integration with respect to
$\theta_M$ in Eqs.~(\ref{eq:order1}) and (\ref{eq:order2}). Owing to
the projection by the integration over $\phi$, the partition function
depends on $\theta_M$ only through the factor
$\mathrm{e}^{\mathrm{i}\mathcal{N}\theta_M}$. Therefore, if
$\mathcal{N}$ takes a non-zero value, the integration with respect to
$\theta_M$ makes zero as a result of the whole average over the phase
factor. Unless external perturbation is introduced,
$\mathcal{P}_{\mathcal{N}\neq0}=0$ should be maintained even in the
thermodynamic limit. In other words, any spontaneous breaking of the
symmetry cannot be described without external perturbation. Thus
even after taking the thermodynamic limit, we have vanishing
$\mathcal{P}_{\mathcal{N}\neq0}$ in the absence of external
perturbation regardless of whether the system lies in the confined or
deconfined phase. This is what Meyer-Ortmanns found in her analysis.

In the presence of small perturbation denoted by $h$, the phase factor
is modified to
$\mathrm{e}^{\mathrm{i}\mathcal{N}\theta_M+2h|M|\cos\theta_M}$. In the
thermodynamic limit, $|M|$ may become as large as proportional to the
volume and then only the stationary point $\theta_M=0$ dominates the
integration with respect to the Polyakov loop. Consequently
 $\mathcal{P_N}$ loses any $\mathcal{N}$ dependence to lead to
$\mathcal{P_N}\neq0$ for arbitrary $\mathcal{N}$. In other
words, the regained center symmetry is always broken and the partition
function is reduced into that in the grand canonical ensemble in which
the center symmetry is explicitly broken whenever small perturbation
is introduced in order to realize the spontaneous symmetry
breaking. Thus the situation is sharply contrast to Meyer-Ortmanns'
analysis. Once external perturbation is applied, the canonical
description resolves itself into the grand canonical description. If
we deal carefully with the thermodynamic limit and try to retain the
canonical description, even the spontaneous breaking of the center
symmetry becomes unavailable. In any case, the state of affairs is
somewhat of antinomy.

In order to gain a deeper insight, we will deal with a toy model in
which the Polyakov loop dynamics is emulated by the spin system in
the next subsection.


\subsection{A toy model}
In this subsection we investigate the thermodynamic limit of the
canonical ensemble by means of the Ising model as a toy tool. The
argument of the center symmetry tells us that the Polyakov loop
dynamics can be effectively described by the Ising model with the Z(2)
symmetry for the SU(2) gauge theory and by the Potts model with the
Z(3) symmetry for the SU(3) gauge theory. In fact, such an effective
description has been established also in the numerical
study~\cite{sve97}.

For simplicity, we will focus here on the Ising model with the Z(2)
symmetry, though the discussion is easily extended to the case of the
Potts model. The counterpart for the order parameter~(\ref{eq:order1})
can be immediately expressed in terms of the spin variables as
\begin{equation}
 \mathcal{P}_{-1}^{\text{Ising}}=\frac{\sum_{\langle s\rangle}
  \mathrm{e}^{-J\sum s_n s_m}\frac{1}{2}\bigl(\mathrm{e}^{H\sum s_n}
  -\mathrm{e}^{-H\sum s_n}\bigr)}{\sum_{\langle s\rangle}
  \mathrm{e}^{-J\sum s_n s_m+H\sum s_n}}
\label{eq:order1sp}
\end{equation}
with the exchange interaction $J$ between the nearest neighbor sites
under a magnetic field $H$. Also the counterpart for the order
parameter~(\ref{eq:order2}) can be written as
\begin{equation}
 \langle s\rangle_{\text{CE}}^{\text{Ising}}=\frac{1}{V}\frac{
  \sum_{\langle s\rangle}\mathrm{e}^{-J\sum s_n s_m}\frac{1}{2}\bigl(
  \mathrm{e}^{H\sum s_n}+\mathrm{e}^{-H\sum s_n}\bigr)\,\sum s_n}
  {\sum_{\langle s\rangle}\mathrm{e}^{-J\sum s_n s_m}\frac{1}{2}\bigl(
  \mathrm{e}^{H\sum s_n}+\mathrm{e}^{-H\sum s_n}\bigr)}.
\label{eq:order2sp}
\end{equation}

For the purpose of investigating what happens in the thermodynamic
limit, it will be sufficient to consider the simplest case in which
the exchange interaction is absent ($J=0$). Then the order parameters
given by Eqs.~(\ref{eq:order1sp}) and (\ref{eq:order2sp}) can be
exactly evaluated, namely, they are trivially zero. The naive
expectation is that these order parameters remain zero while the
magnitude of the exchange interaction is small, regardless of the
presence of an external magnetic field $H\neq0$. They will come to
take a finite value when the strength of the interaction grows large
enough to bring about the dynamical symmetry breaking. Actually,
however, this naive expectation is not realized at all. We must follow
the standard procedure to describe the spontaneous symmetry breaking;
we add an infinitesimal external field denoted by $h$ and then take
the thermodynamic limit ($V\to\infty$) and finally turn off the
external field ($h\to0$). The order parameters under an infinitesimal
external field $h$ is calculated for $J=0$ as
\begin{align}
 \mathcal{P}_{-1}^{\text{Ising}} &=\frac{\sum_{\langle s\rangle}
  \frac{1}{2}\bigl(\mathrm{e}^{(H+h)\sum s_n}-\mathrm{e}^{-(H-h)
  \sum s_n}\bigr)}{\sum_{\langle s\rangle}
  \mathrm{e}^{(H+h)\sum s_n}}\notag\\
 &= \frac{\bigl(2\cosh(H+h)\bigr)^V-(2\cosh(H-h)\bigr)^V}{2\bigl(
  2\cosh(H+h)\bigr)^V}\notag\\
 &\to \frac{1}{2}\qquad (V\to\infty,\quad H\gg h>0).
\label{eq:thermo1}
\end{align}
We can draw the similar result for another candidate, that is,
\begin{align}
 \langle s\rangle_{\text{CE}}^{\text{Ising}} &=\frac{1}{V}\frac{
  \sum_{\langle s\rangle}\frac{1}{2}\bigl(\mathrm{e}^{(H+h)\sum s_n}
  +\mathrm{e}^{-(H-h)\sum s_n}\bigr)\sum s_n}{\sum_{\langle s\rangle}
  \frac{1}{2}\bigl(\mathrm{e}^{(H+h)\sum s_n}+\mathrm{e}^{-(H-h)\sum
  s_n}\bigr)}\notag\\
 &=\frac{\bigl(2\cosh(H+h)\bigr)^{V-1}2\sinh(H+h)-\bigl(2\cosh(H-h)
  \bigr)^{V-1}2\sinh(H-h)}{\bigl(2\cosh(H+h)\bigr)^V+\bigl(2\cosh(H-h)
  \bigr)^V}\notag\\
 &\to\tanh(H+h)\qquad (V\to\infty,\quad H\gg h>0).
\label{eq:thermo2}
\end{align}
The above results of Eqs.~(\ref{eq:thermo1}) and (\ref{eq:thermo2})
clearly signify that the would-be order parameters given by
Eqs.~(\ref{eq:order1sp}) and (\ref{eq:order2sp}) are to be
disqualified under the thermodynamic limit because they always remain
finite. In other words, any expectation value calculated in the
canonical ensemble in which the center symmetry is seemingly recovered
would be reduced into that calculated in the grand canonical ensemble
in which the symmetry is explicitly broken. The point is that the
contribution among the projective superposition with respect to $\phi$
survives only when it is parallel to the direction of the external
magnetic field. As a result, the projection into the canonical
ensemble is diminished only to result in the same description as in
the grand canonical ensemble. We can intuitively understand the
situation in the following way: In the thermodynamic limit, the
projective superposition in Eq.~(\ref{eq:order2sp}) for example
becomes
\begin{equation}
 \mathrm{e}^{H\sum s_n}+\mathrm{e}^{-H\sum s_n}\to
  \mathrm{e}^{H|\sum s_n|},
\label{eq:projection}
\end{equation}
which follows that the spin configurations for which
$\mathrm{e}^{H\sum s_n}$ dominates should be definitely separated from
the spin configuration for which $\mathrm{e}^{-H\sum s_n}$ does, that
is, the system becomes non-ergodic. Which term survives in the
thermodynamic limit depends on which direction an external field is
introduced in.

The same mechanism  makes the canonical ensemble in QCD be unstable
under the thermodynamic limit. For the configuration of the Polyakov
loop with a macroscopic order, $M=\sum\mathrm{tr}L(\vec{x})=|M|\,
\mathrm{e}^{\mathrm{i}\theta_M}\sim\mathrm{O}(V)\to\infty$, the
integration with respect to $\phi$ in Eq.~(\ref{eq:integ_phi}) becomes
\begin{align}
 & \int_0^{2\pi}\frac{\mathrm{d}\phi}{2\pi}\,\mathrm{e}^{-\mathrm{i}
  n_q\phi}\cdot\mathrm{e}^{H(\mathrm{e}^{\mathrm{i}\phi}M+
  \mathrm{e}^{-\mathrm{i}\phi}M^\dagger)}
 = \int_0^{2\pi}\frac{\mathrm{d}\phi}{2\pi}\,\mathrm{e}^{-\mathrm{i}
  n_q\phi}\cdot\mathrm{e}^{2H|M|\cos(\phi+\theta_M)}\notag\\
 \simeq& \frac{\sqrt{2\pi}\,\mathrm{e}^{-\mathrm{i}n_q\phi_0}\cdot
  \mathrm{e}^{2H|M|\cos(\phi_0+\theta_M)}}{2\pi\bigl|-2H|M|
  \cos(\phi_0+\theta_M)\bigr|^{\frac{1}{2}}}
  \Biggl|_{\phi_0=-\theta_M}\qquad (|M|\sim\mathrm{O}(V)\to\infty).
\label{eq:saddle}
\end{align}
In the last line of the above equations only the leading order of the
saddle-point approximation ($2H\cos(\phi_0+\theta_M)$ is maximized at
$\phi_0=-\theta_M$) is left, which becomes exact when $|M|$ goes to
infinity. The meaning of the saddle-point approximation in
Eq.~(\ref{eq:saddle}) is absolutely the same as what
Eq.~(\ref{eq:projection}) means. Then infinitesimal perturbation like
the $h$ field in Eqs.~(\ref{eq:thermo1}) and (\ref{eq:thermo2})
forces only specific $\theta_M$ to be favored in the thermodynamic
limit. Thus neither DeTar--McLerran's order parameter nor the Polyakov
loop expectation value in the canonical ensemble would serve as a
proper order parameter for the deconfinement transition.


\section{IDEA TO OVERCOME THE PROBLEM}
\label{sec:idea}


\subsection{Prospect as order parameters}
So far we have clarified why DeTar--McLerran's order parameter does
not work as expected under the thermodynamic limit. Nevertheless, we
have prospects of making use of DeTar--McLerran's order parameter to
identify the deconfinement transition. The idea is quite simple. The
spin-system counterpart for DeTar--McLerran's order parameter given
in Eq.~(\ref{eq:order1sp}) can be equivalently expressed in the
following form:
\begin{equation}
 \mathcal{P}_{-1}^{\text{Ising}}=\frac{\frac{1}{2}\langle
  \mathrm{e}^{H\sum s_n}-\mathrm{e}^{-H\sum s_n}\rangle_J}{\langle
  \mathrm{e}^{H\sum s_n}\rangle_J},
\label{eq:order1sp2}
\end{equation}
where $\langle\cdots\rangle_J$ denotes the ensemble average with the
action $-J\sum s_n s_m$. Let us imagine that we compute the order
parameter of Eq.~(\ref{eq:order1sp2}) by the method of the Monte-Carlo
simulation. First we generate the ensemble of configurations with the
probability specified by the action $-J\sum s_n s_m$, which has both
the symmetric phase with random alignments and the broken phase with a
spontaneous magnetization. Next we calculate the ensemble average with
those generated configurations. In the symmetric phase, the numerator
of Eq.~(\ref{eq:order1sp2}) should vanish, while it take a finite
value in the broken phase. Hence it is apparent that the would-be
order parameter given in Eq.~(\ref{eq:order1sp2}) will serve as a
proper order parameter. If we prefer the spin expectation value to
DeTar--McLerran's formulation, it would be enough to rewrite
Eq.~(\ref{eq:order2sp}) into
\begin{equation}
 \langle s\rangle_{\text{CE}}^{\text{Ising}}=\frac{1}{V}\frac{\langle
  \bigl(\mathrm{e}^{H\sum s_n}+\mathrm{e}^{-H\sum s_n}\bigr)\sum s_n
  \rangle_J}{\langle\mathrm{e}^{H\sum s_n}+\mathrm{e}^{-H\sum s_n}
  \rangle_J}.
\label{eq:order2sp2}
\end{equation}
What is important here is that the probability for
configurations is specified by the symmetric part of the action. If we
adopt the whole action, for example
$-J\sum s_n s_m+\ln\cosh(H\sum s_n)$ for Eq.~(\ref{eq:order2sp}), to
generate the configurations, the ensemble of configurations becomes
non-ergodic in the thermodynamic limit and the would-be order
parameters should inevitably fail as demonstrated in
Eqs.~(\ref{eq:thermo1}) and (\ref{eq:thermo2}). In other words, the
ensemble of configurations given by
$-J\sum s_n s_m+\ln\cosh(H\sum s_n)$ is unstable against
configurations with $\sum s_n\sim\mathrm{O}(V)$ because of the
long-ranged nature of the interaction term $\ln\cosh(H\sum s_n)$. As a
result the spin configurations with $\sum s_n>0$ is completely
decoupled from those with $\sum s_n<0$ and the canonical description
is reduced to the grand canonical one.

The above idea is quite simple in itself. In the rest of this
subsection, we will discuss the physical meaning of the above
prescription to generate the ensemble of configurations with the
probability specified by the symmetric part of the action.

Looking back at the argument around Eq.~(\ref{eq:saddle}), let us
reconsider the validity for the saddle-point approximation which is
expected to be exact in the thermodynamic limit. Once $|M|$ is really
as large as of order $\mathrm{O}(V)$, the argument given in
Eq.~(\ref{eq:saddle}) has no suspicious controversy at all. However,
going back to the second line of Eq.~(\ref{eq:CE_QCD}), we realize
that the coefficient in front of $\phi$ is at most $n_q$ which is now
of order $\mathrm{O}(1)$. If we considered the thermodynamic limit
where we take $n_q\sim\mathrm{O}(V)\to\infty$ with the number density
$n_q/V$ fixed, the aforementioned argument of the saddle-point
approximation would become exact and the calculation of the canonical
ensemble is absolutely identical as that in the grand canonical one,
as closely discussed in Ref.~\cite{hag85}. Then the expectation value
of the phase of the Polyakov loop is finite and almost proportional to
the quark number density~\cite{for00}, as seen in Eq.~(\ref{eq:nq}).

As we have already emphasized, the canonical ensemble for the purpose
of characterizing the deconfinement transition is different. What we
should be confronted with is the ensemble where $n_q$ is fixed at so
small number of order $\mathrm{O}(1)$ that the saddle-point
approximation would be inapplicable. So as to realize this situation
an additional condition is needed; $|M|=|\sum\mathrm{tr}L(\vec{x})|$
must be kept of order $\mathrm{O}(1)$. This condition can be achieved
in the form of the ensemble average in Eqs.~(\ref{eq:order1sp2}) and
(\ref{eq:order2sp2}) as numerically demonstrated in the next
subsection.

We shall summarize our findings and assertions here again.
\begin{itemize}
\item Even when the quark number is fixed in the canonical
formulation, it is the quark number \textit{density} that is actually
fixed under the thermodynamic limit because the system becomes
unstable against configurations with a macroscopic order. Thus any
fluctuation of the quark number of order $\mathrm{O}(1)$ is allowed
even in the canonical description, though it is unintended. As a
result the notion of confinement becomes obscured.
\item The idea is simple: If we completely get rid of the fluctuation
of the quark number, which is negligible in the thermodynamic limit
but responsible for screening the Polyakov loop, the canonical
ensemble with respect to the quark number is expected to recover its
meaning. Then DeTar--McLerran's order parameter and also the Polyakov
loop should serve as order parameters for the deconfinement
transition.
\item The fluctuation of the quark number is induced by the infinitely
non-local interaction arising from the projective superposition. Such
fluctuation can be excluded by taking the ensemble average with
configurations generated by the symmetric part of the action. The
asymmetric part of the action is regarded as included in the operator
part to be averaged over.
\end{itemize}


\subsection{Numerical tests}
\begin{figure}
\includegraphics[width=7cm]{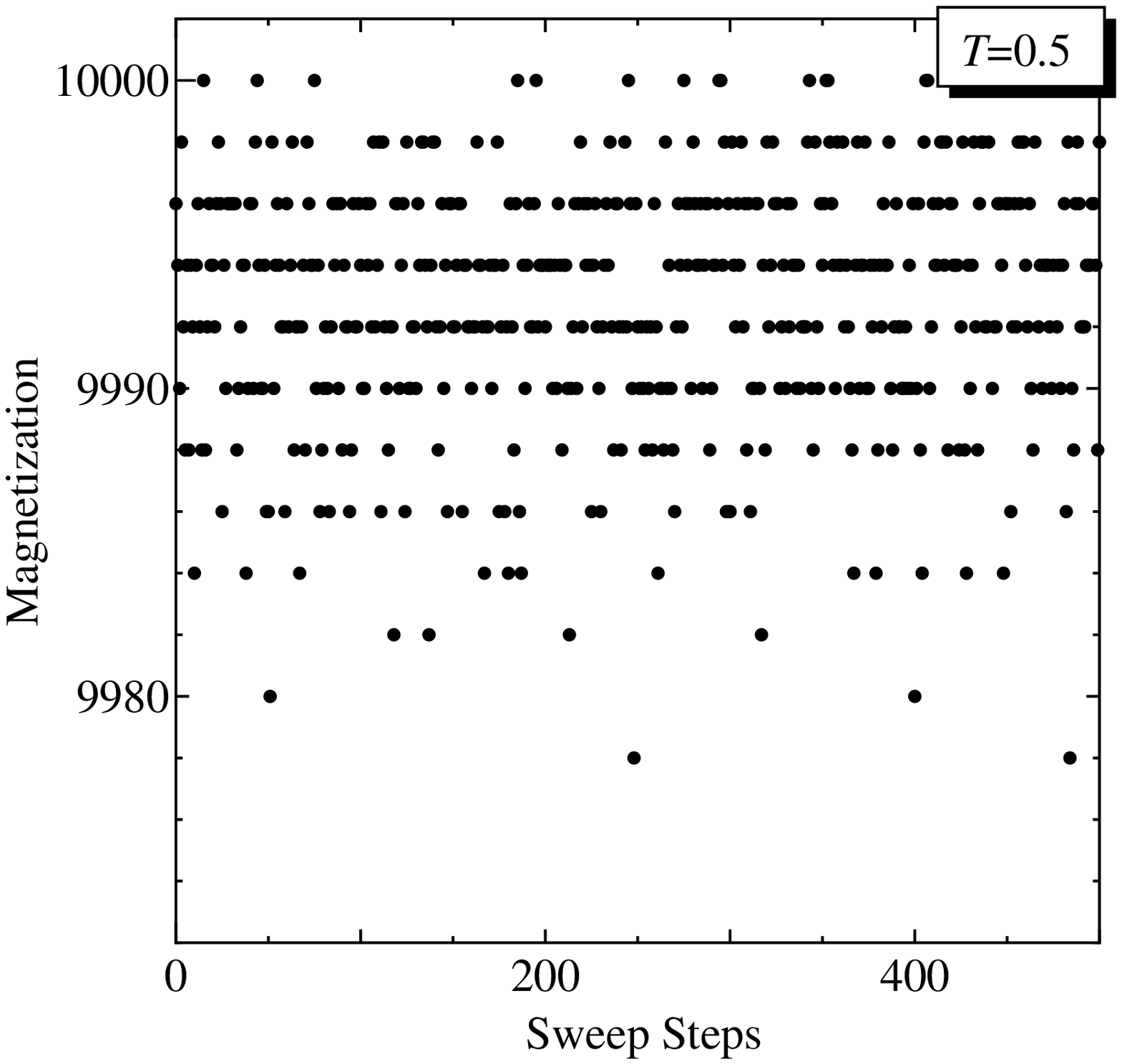}\hspace{1cm}
\includegraphics[width=7cm]{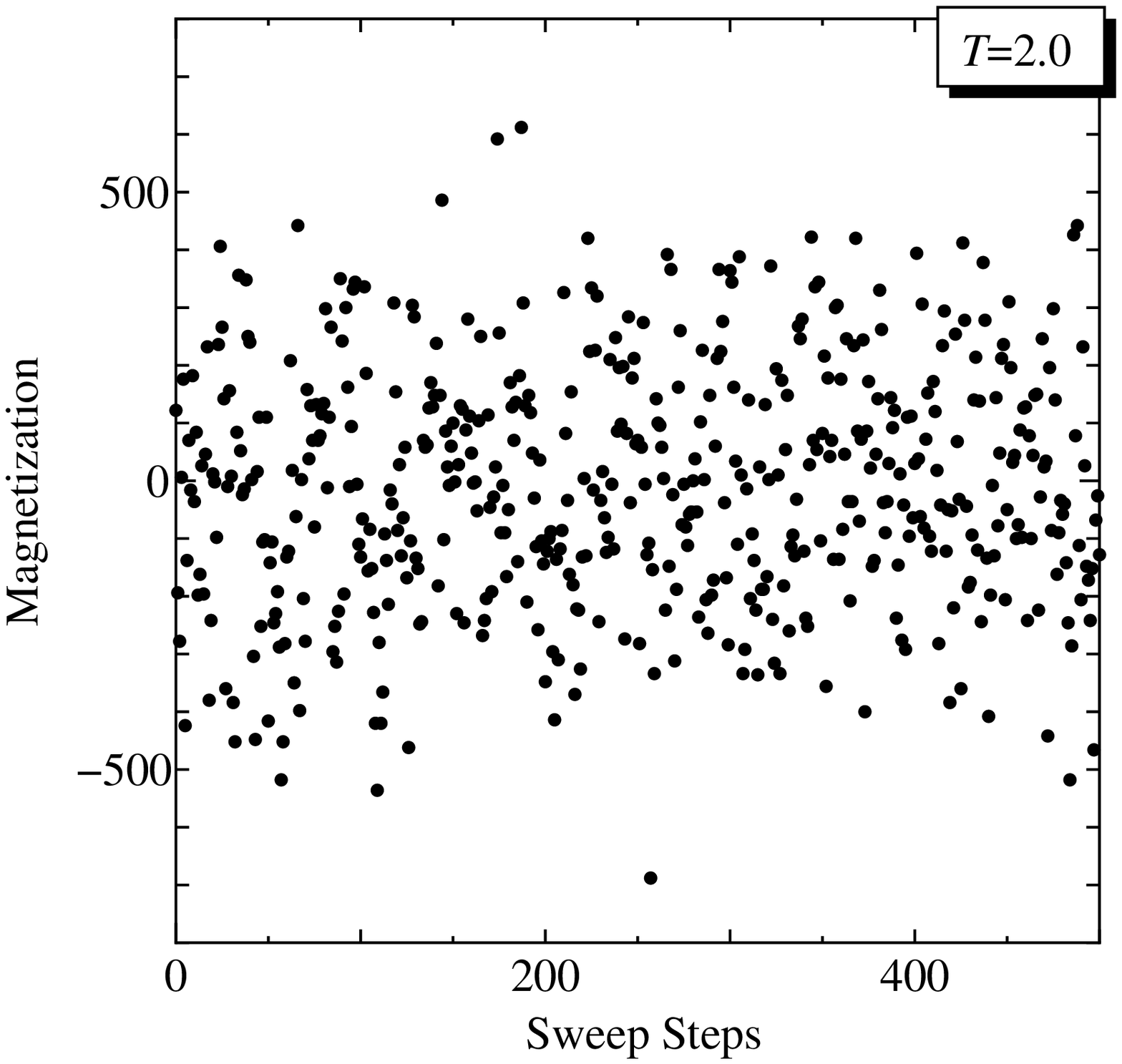}
\caption{Magnetization at each sweep step after the thermalization
achieved by 10000 times sweeps. The left figure is for $T=0.5$ in the
ordered phase and the right for $T=2.0$ in the disordered phase.}
\label{fig:sweep}
\end{figure}

By using the Monte-Carlo simulation, we numerically calculate
Eq.~(\ref{eq:order2sp2}) which can be comparable with an ordinary
magnetization in the presence of an external magnetic field. The
lattice size is chosen as $100\times100$ and we adopt the standard
Metropolis algorithm. The strengths of the exchange interaction and
the magnetic field are $J=0.5/T$ and $H=0.1/T$ respectively, where $T$
is the temperature. The thermalization of the spin system is well
achieved after 10000 times sweeps, as can be confirmed by the
magnetization distributions shown in Fig.~\ref{fig:sweep} for $T=0.5$
and $T=2.0$. It should be noted that the magnetization is typically of
order $\mathrm{O}(V)$ in the ordered phase ($T=0.5$) and of order
$\mathrm{O}(1)$ in the disordered phase ($T=2.0$).

\begin{figure}
\includegraphics[width=7cm]{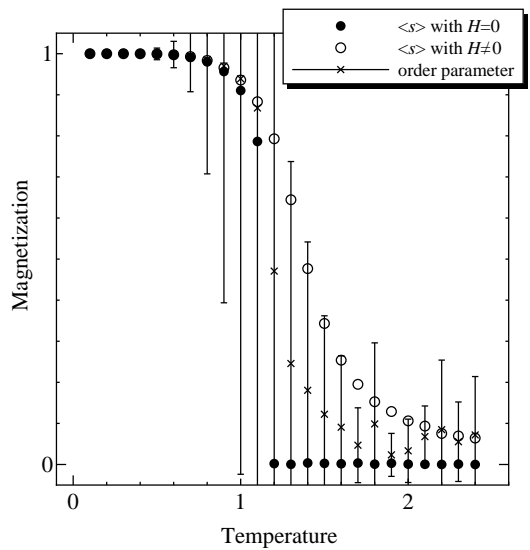}
\caption{Behaviors of the magnetizations and the order parameter as
functions of the temperature.}
\label{fig:magnet}
\end{figure}

The resultant $\langle s\rangle_{\text{CE}}^{\text{Ising}}$ is
presented in Fig.~\ref{fig:magnet}. Filled circles stand for the
spontaneous magnetization in the case of $H=0$ and indicate that the
spin system goes through the second-order phase transition at
$T_{\text{c}}=2.27J=1.13$ (theoretical value). Blank circles represent
the magnetization in the presence of the magnetic field $H\neq0$ and
show a smooth crossover as the temperature raises. We plot
$\langle s\rangle_{\text{CE}}^{\text{Ising}}$ by crosses with
error-bars. Although we take 10000 samples by each 10 intervals, the
statistical error is still large because the term
$\mathrm{e}^{H\sum s_n}+\mathrm{e}^{-H\sum s_n}$ in both the numerator
and the denominator of Eq.~(\ref{eq:order2sp2}) can become so large
that the operator to be taken the ensemble average of must furiously
fluctuate to lead to large statistical errors. In a sense we can say
that this problem of slow convergence arises from the same origin as
that of the sign problem for simulations at finite density, in
particular, the situation is alike the Glasgow method in lattice QCD
at finite density. This parallelism to the sign problem will be more
apparent for the order parameter given by Eq.~(\ref{eq:order1sp2}) or
Eq.~(\ref{eq:order1}). As discussed in Sec.~\ref{sec:limit} the
canonical partition function can vanish due to the average over the
phase factor coming from the Dirac determinant.

The results shown in Fig.~\ref{fig:magnet} seem to suggest that
the would-be order parameter
$\langle s\rangle_{\text{CE}}^{\text{Ising}}$ should work as
expected. Unfortunately, at this stage, we cannot address any stronger
statement due to large statistical errors. In order to improve the
accuracy, some essential ingenuity such as a reweighting
method~\cite{fod01} are needed rather than simple accumulation of many
samples. That is beyond the scope of the present paper.


\subsection{Construction of the order parameter in QCD}
It is easy to write down the QCD counterparts for the order parameters
given by Eqs.~(\ref{eq:order1sp2}) and (\ref{eq:order2sp2}) in the
spin system. Here we present the expressions only for
DeTar--McLerran's order parameter defined by Eq.~(\ref{eq:detar}) and
the Polyakov loop expectation value given by Eq.~(\ref{eq:pol}).
The formulation can be straightforwardly extended to general cases.

The quark contribution from the Dirac determinant can be expressed as
the effective action in the following form;
\begin{equation}
 \det D[U]=\mathrm{e}^{-S_{\text{eff}}[U,U_4(x_4=N_\tau)]},
\end{equation}
where the link variable $U_4(x_4=N_\tau)$ receiving a modification
under the center transformation is explicitly written for convenience.
We can decompose the effective action according to the triality, that
is,
\begin{equation}
 S_{\text{eff}}^{(k)}[U]=\frac{1}{3}\sum_{n=0,\pm1}\,
  \mathrm{e}^{-\mathrm{i}2\pi kn/3}\,S_{\text{eff}}[U,z_0^n
  U_4(x_4=N_\tau)]\qquad (k=0,\pm1)
\end{equation}
with $z_0=\mathrm{e}^{\mathrm{i}2\pi/3}\in\mathrm{Z}(3)$. Among the
decomposed parts, all the contributions with non-trivial triality
($k=\pm1$) are bound to be included in the operator to be averaged
over. Therefore the would-be order parameter given by
Eq.~(\ref{eq:detar}) can be written for $\mathcal{N}=1$ as
\begin{equation}
 \tilde{\mathcal{P}}_1=\frac{\frac{1}{3}\Bigl\langle\mathrm{e}^{
  -S_{\text{eff}}^{(1)}-S_{\text{eff}}^{(-1)}}+z_0^{-1}
  \mathrm{e}^{-z_0 S_{\text{eff}}^{(1)}-z_0^{-1}
  S_{\text{eff}}^{(-1)}}+z_0\,\mathrm{e}^{-z_0^{-1}
  S_{\text{eff}}^{(1)}-z_0 S_{\text{eff}}^{(-1)}}\Bigr
  \rangle_{\text{sym}}}
  {\Bigl\langle\mathrm{e}^{-S_{\text{eff}}^{(1)}-
  S_{\text{eff}}^{(-1)}}\Bigr\rangle_{\text{sym}}}.
\end{equation}
The ensemble average is to be taken with the probability weight
specified by the symmetric part of the action, \textit{i.e.},
$-S_{\text{G}}[U]-S_{\text{eff}}^{(0)}[U]$. In the same way, the
expression for Eq.~(\ref{eq:pol}) can be written as
\begin{equation}
 \langle\widetilde{\mathrm{tr}L(\vec{x})}\rangle_{\text{CE}}=
  \frac{\Bigl\langle\bigl(\mathrm{e}^{-S_{\text{eff}}^{(1)}-
  S_{\text{eff}}^{(-1)}}+\mathrm{e}^{-z_0 S_{\text{eff}}^{(1)}-
  z_0^{-1}S_{\text{eff}}^{(-1)}}+\mathrm{e}^{-z_0^{-1}
  S_{\text{eff}}^{(1)}-z_0 S_{\text{eff}}^{(-1)}}\bigr)\,\mathrm{tr}
  L(\vec{x})\Bigr\rangle_{\text{sym}}}
  {\Bigl\langle\mathrm{e}^{-S_{\text{eff}}^{(1)}-
  S_{\text{eff}}^{(-1)}}+\mathrm{e}^{-z_0 S_{\text{eff}}^{(1)}-
  z_0^{-1}S_{\text{eff}}^{(-1)}}+\mathrm{e}^{-z_0^{-1}
  S_{\text{eff}}^{(1)}-z_0 S_{\text{eff}}^{(-1)}}\Bigr
  \rangle_{\text{sym}}}.
\end{equation}


\section{CONCLUDING REMARKS}
\label{sec:remarks}
In order to identify the deconfinement phase transition we employed
and elaborated the idea of the canonical ensemble with respect to the
quark number or triality. We clarify why the canonical ensemble would
be reduced into the grand canonical ensemble eventually in the
thermodynamic limit. In order to overcome the problem of the
thermodynamic limit, we propose a prescription to compute the ensemble
average. The idea is tested by means of an effective model in terms
of spin variables in the presence of the magnetic field. The results
seem prosperous but turn out to suffer from the severe sign problem.
Although our definition of the order parameter for the deconfinement
transition can be applied to the Monte-Carlo simulation on the
lattice, it is inevitably necessary to get over the sign problem
inherent in the lattice QCD at finite density. We find it interesting
that two serious questions -- one is the criterion for the
deconfinement transition at finite temperature, and the other is the
sign problem in the lattice QCD at finite density -- are closely
related to each others.

\begin{acknowledgments}
The author, who is supported by Research Fellowships of the Japan
Society for the Promotion of Science for Young Scientists, thanks
Professor J.~Polonyi for stimulating discussions.
\end{acknowledgments}

\end{document}